\documentclass[twocolumn]{aastex631}
\usepackage{amsmath}
\usepackage{placeins}

\begin{document}

\title{The Prompt Emission of Gamma-Ray Bursts from Population III Stars}

\author[0009-0008-1678-2787]{Nathan Walker}
\affil{Department of Physics, Oregon State University, 301 Weniger Hall, Corvallis, OR 97331, USA}

\author[0000-0002-9190-662X]{Davide Lazzati}
\affil{Department of Physics, Oregon State University, 301 Weniger Hall, Corvallis, OR 97331, USA}

\author[0000-0002-4299-2517]{Tyler M. Parsotan}
\affil{Astrophysics Science Division, NASA Goddard Space Flight Center, Greenbelt, MD 20771, USA}

\author[0000-0002-3635-5677]{Rosalba Perna}
\affil{Department of Physics and Astronomy, Stony Brook University, Stony Brook, NY 11794-3800, USA}

\author[0000-0002-8171-8596]{Matteo Cantiello}
\affiliation{Center for Computational Astrophysics, Flatiron Institute, New York, NY 10010, USA}
\affiliation{Department of Astrophysical Sciences, Princeton University, Princeton, NJ 08544, USA}

\author[0009-0004-5685-6155]{Jake B. Hassan}
\affil{Department of Physics and Astronomy, Stony Brook University, Stony Brook, NY 11794-3800, USA}

\author[0000-0003-3846-257X]{Brian Morsony}
\affil{Department of Physics and Physical Sciences, Stanislaus State University, Turlock, CA, 95382}

\begin{abstract}
Population (Pop) III stars, the first generation of stars in the Universe, have eluded direct detection for decades. The advent of the James Webb Space Telescope, with its unprecedented sensitivity to high-redshift sources, has renewed interest in the prospects for identifying these primordial stars. While previous studies have explored the direct detection of Pop III stars through gravitational lensing, here we investigate an alternative approach: identifying Pop III stars through gamma-ray bursts (GRBs) produced by the deaths of massive Pop III progenitors. We generate source-frame mock observables and find that, although Pop III and Pop I GRBs share many common properties, several observables can potentially distinguish between the two populations. In particular, light-curve timescales, isotropic-equivalent energies, peak energies, and the locations of GRBs in empirical prompt-emission correlations depend sensitively on the mass and explosion energy of the progenitor. We find that Pop III GRBs tend to exhibit longer characteristic timescales and higher isotropic-equivalent peak energies than their Pop I counterparts. Finally, we show that Pop III GRBs may provide a promising origin for the population of soft X-ray prompt emission recently detected by the Einstein Probe, offering a complementary avenue for probing the first generations of stars.

\end{abstract}

\section{Introduction}
Observational inferences of current stellar populations support the existence of completely metal-free stars that existed in the early universe. While these Population III stars have yet to be detected, their formation at a time when the matter in the universe consisted almost entirely of hydrogen and helium gives them an instrumental role in shaping the metallicity of future generations of Population I and II stars \citep{abel_formation_2002,bromm_formation_2002,turk_formation_2009}. Recent studies on the observability of both rotating \citep{hassan_spectral_2025} and non-rotating \citep{larkin_characterization_2022} Population III stars suggest the possibility of observing  the most massive of these stars through gravitational lensing, which can result in  magnifications of $\mu \lesssim 10^3$. However, in the absence of direct detections of Population III stars though these, or other, methods, it is critical to assess other potential avenues of identifying Population III stars.  

While direct observations of Population III stars are challenging due to their high redshift, one possible avenue of detection is through associated gamma-ray bursts (GRBs). Canonical long-duration GRBs (LGRBs), originating from Population I and II stars, are thought to be produced during the core-collapse of massive stars \citep{galama_unusual_1998,costa_discovery_1997,van_paradijs_transient_1997,metzger_spectral_1997}. After the formation of a rotating compact object during core-collapse, the continued in-fall of stellar material can form an accretion disk around the remnant and power bi-polar jets \citep{gottlieb_black_2022,tchekhovskoy_efficient_2011}. These jets, if powerful enough to bore their way through the collapsing stellar envelope, can produce large amounts of observable high-energy gamma-rays \citep{macfadyen_collapsars_1999}. 

The impact a stellar progenitor has on a GRB can begin at the earliest stages of the stars evolution, which for Population III stars involves the gravitational collapse of molecular hydrogen and helium in the very early universe. Initial considerations based upon this fact indicated that Population III stars could have masses of up to the local Jeans mass of $\sim1000 M_{\odot}$ \citep{abel_formation_2002,bromm_formation_2002}. However, subsequent work has suggested that a more reasonable upper limit brought upon by fragmentation of the primordial gas is $\sim100 M_{\odot}$ \citep{mckee_formation_2008,hosokawa_protostellar_2011,wollenberg_formation_2020, jaura_trapping_2022, hassan_spectral_2025}. Even with this more conservative upper limit, Population III stars are still expected to be significantly more massive than their Population I counterparts. Additionally, the chemical makeup of Population III stars can lead to high rotation \citep{hassan_spectral_2025}, which can have a dramatic impact on subsequent jets. 

After a jet is launched by the central engine it must breakout of the stellar envelope. As the structure of a stellar progenitor has been shown to impact the structure of GRBs \citep{morsony_origin_2010}, this fact, together with the aforementioned discussion on the formation of Population III stars, provides a compelling case that, if detectable, Population III GRBs can be distinguished from Population I and II GRBs. As such, there have been several studies on the production of GRBs from Population III stars. These include investigations into their break-out criteria \citep{nagakura_population_2012}, and on the efficiency of producing such GRBs \citep{morales-rivera_efficiency_2025}. The former is of greatest relevance to this work, as it lays the groundwork for hydrodynamical simulations of Population III GRBs, including what is, to the best of our knowledge, the only instance of a GRB simulation that calculates the jet luminosity self-consistently from the accretion of a stellar progenitor onto a compact central engine.

Here we present the results of the first simulations of the prompt emission from Population III GRBs. Building off previous work \citep{perna2018,hassan_spectral_2025}, we inject a relativistic jet into a  Population III stellar progenitor via a self-consistent procedure and simulate the jet until it becomes transparent to radiation. Then, we use the radiative transfer software MCRaT \citep{lazzati_monte_2016} to produce predictions of GRB prompt emission. Because the implementation of a self-consistent jet can in principle lead to substantially different dynamics than what has been observed in past work, we carry out this procedure for a Population I  stellar progenitor as well.

This paper is organized as follows: in Section 2 we discuss the stellar evolution of our progenitor stars. In Section 3 we describe our hydrodynamical simulations, including the numerical setup and new techniques we use to launch relativistic jets. In Section 4 we present the emission of LGRBs, and in Section 5 we discuss implications and future work. Data from this work, including the hydrodynamical initial conditions and the results of our radiative transfer calculations, is made available at \url{https://doi.org/10.5281/zenodo.23045935}.

\section{Stellar Evolution}

 We compute the properties of stellar progenitors using \texttt{MESA} version 24.08.1 \citep{paxton_modules_2010,paxton_modules_2013,paxton_modules_2018,paxton_modules_2015,paxton_modules_2019}. \texttt{MESA} is an open-source, community-driven project that solves the 1D equations of stellar evolution.  While \texttt{MESA} is 1D in nature, it models stellar rotation using the shellular approximation \citep{Zahn1992, MaederZahn1998}, with angular momentum and chemical transport treated as diffusive processes following standard 1D prescriptions \citep[e.g.,][]{Heger2000,Heger2005,paxton_modules_2013,Yoon2012}. 

The Population III stellar progenitor we use in this work builds off the models of  \citet{hassan_spectral_2025}, which computed the stellar evolution of rotating Population III stars until the onset of Helium-burning. Here, we continue the stellar evolution for one representative model until the pre-supernova stage. The model we choose is a pre-main-sequence star with initial mass $M=37 M_\odot$, metallicity $Z=0$, initial composition $X=0.76$, $Y=0.24$ \citep{Yoon2012},
a core temperature $T_c=5\times 10^5$~K \citep{larkin_characterization_2022,hassan_spectral_2025}, and
with a surface velocity of  $v_{\mathrm{s}} = 0.6\,v_{\mathrm{crit}}$, where
\begin{equation}
    v_{\mathrm{crit}} = \sqrt{\left(1 - \frac{L}{L_{\mathrm{edd}}}\right) \frac{GM}{R}}
\end{equation}
is defined in terms of the Eddington Luminosity $L_{\mathrm{edd}} = 4\pi\, c\,GM/\kappa$, where $\kappa$ is the opacity. Additionally, we adopt the Skye equation of state \citep{Jermyn2021}.

\begin{figure}
    \centering
    \includegraphics[width=\columnwidth]{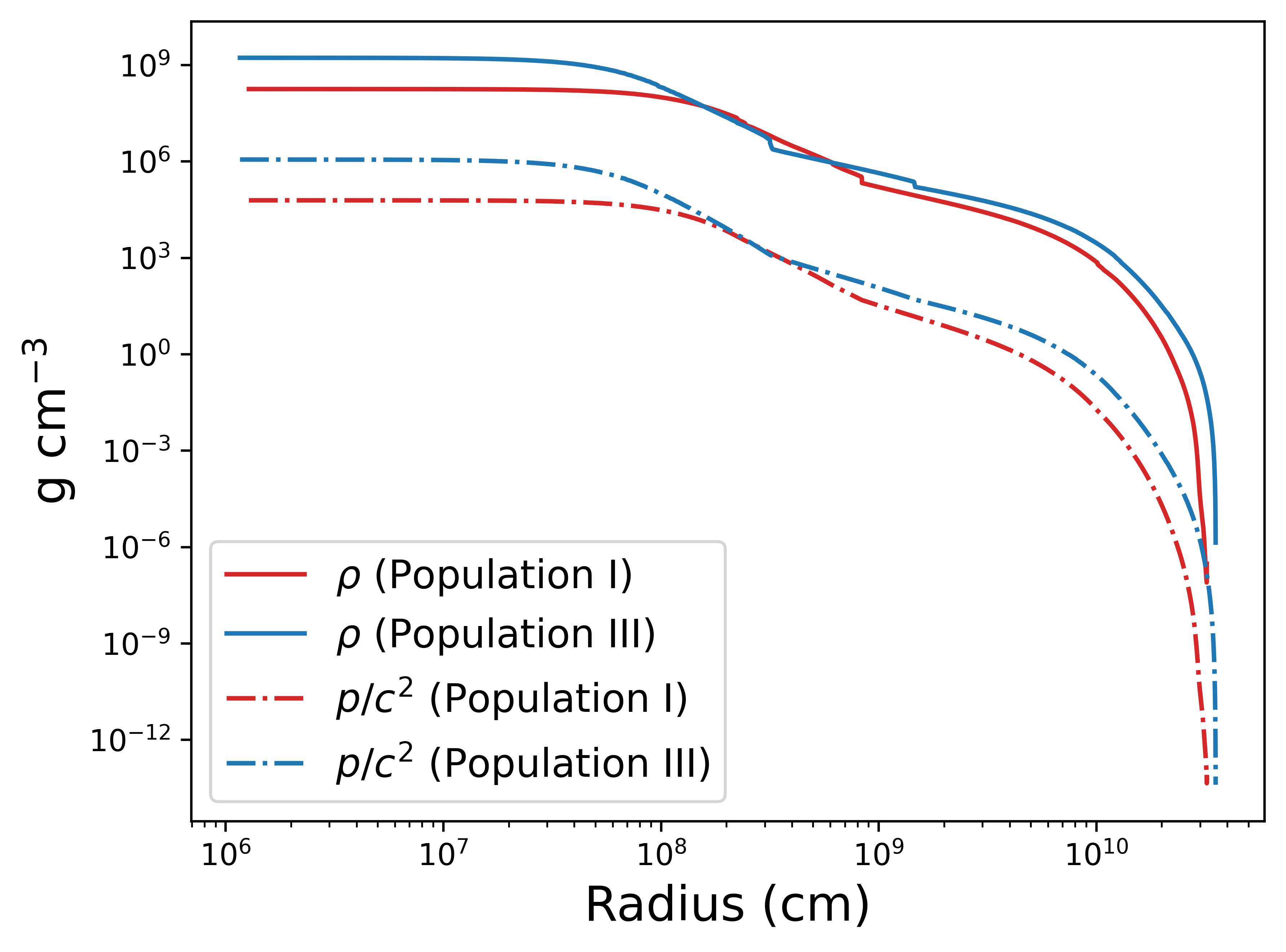}
    \caption{Stellar density (solid line) and pressure (dash-dotted line) as a function of radius for the Population I (red) and Population III (blue) progenitors used in this work. }
    \label{fig:initial_density}
\end{figure}

As a reference comparison of a 'canonical' GRB, we also study jet propagation in a Population~I stellar progenitor with $v_{\mathrm s} = 0.5 v_{\mathrm{crit}}$ and initial mass $M=20 M_{\odot}$, taken from the pre-supernova models of \citet{perna_fate_2014}. This Population I progenitor was chosen to have a similar rotational velocity. Density and pressure profiles of both models are shown in Figure \ref{fig:initial_density}.

We note that while the stellar evolution of Population~III stars is a topic of considerable interest, this paper is focused on jet dynamics that occur in the stellar interior and at distances far away from the central engine, and we refer to previous references for more details on our progenitors stellar evolution.

\section{Hydrodynamic simulations of jet propagation}

We use the relativistic hydrodynamics (RHD) module of the \texttt{PLUTO} hydrodynamical code \citep{mignone_pluto_2007} to inject and propagate jets into stellar progenitors. Both the initial conditions and the jet injection parameters are derived from the stellar models discussed in the previous section. Furthermore, we build off the foundational work \cite{nagakura_jet_2011} by injecting a jet self-consistently into the computational domain, as opposed to adopting either a constant or analytic jet luminosity. To date, this is to the best of our knowledge the only other work that has injected a jet based off the real-time conditions of a hydrodynamical system. By continuing this effort to base jet injection on real-time dynamics, we allow properties of the stellar progenitor to impact the jet evolution and, eventually, its emission.

\subsection{Numerical Setup}

\subsubsection{Initial Conditions}

We carry out hydrodynamical simulations in 2D spherical coordinates for both Population I and Population III stellar progenitors. In both cases the computational domain extends from $\theta=0^{\circ}$ to $\theta=90^{\circ}$ relative to the jet axis, and out to $10^{14}$ cm radially to allow for a long-lived central engine. The inner radius is $r_{\mathrm{in}}=2.3 \times 10^{9}$ cm and $r_{\mathrm{in}} = 3.4 \times 10^{9}$ cm for the Population I and III progenitors, respectively, resulting in a radial resolution of $\Delta r \sim 2 \times 10^{8}$ cm for both simulations. $r_{\mathrm{in}}$ is chosen based on the black hole mass after core collapse (see below).

Initial conditions for the stellar interior are obtained from \texttt{MESA} output as described in the previous section (see Figure \ref{fig:initial_density}). We then surround each progenitor with the density profile
\begin{equation}
    \rho_{\mathrm w} = \frac{\rho_{0}}{r^2}
\end{equation}
where $\rho_{0}=8.38 \times 10^{13} \, \mathrm{g} \, \mathrm{cm}^{-3}$, and was chosen to ensure numerical stability. While the density exterior to the star is in principle due to ejected wind, values derived from this wind are extremely small compared to the density of the star. Adopting these values in simulations leads to numerical errors, including plug instabilities\footnote[1]{Plug instabilities (e.g. \citealt{lazzati_unifying_2012}) frequently appear in 2D axisymmetric simulations of jets. The presence of a symmetry axis creates an obstacle for material in the jet head, which can result in a low-velocity, high density plug that disrupts the jet. The presence of this instability is highly sensitive to initial conditions, and is here minimal enough to not affect the large-scale jet dynamics. } capable of disrupting the jet. On the other hand, while the value we adopt here is large, it is not enough to introduce steep gradients near the stellar surface. Additionally, the amount of energy injected by the jet is large enough to be unaffected by the wind close to the jet axis. It is worth noting, however, that wind parameters closer to the stellar wind will be necessary to probe prompt properties of GRBs farther off-axis than we consider here.

% for the Population I model and $\rho_{0}=9.42 \times 10^{16} \, \mathrm{g} \, \mathrm{cm}^{-3}$ for the Population III model

\begin{figure*}
    \centering
    \includegraphics[width=1.0\textwidth]{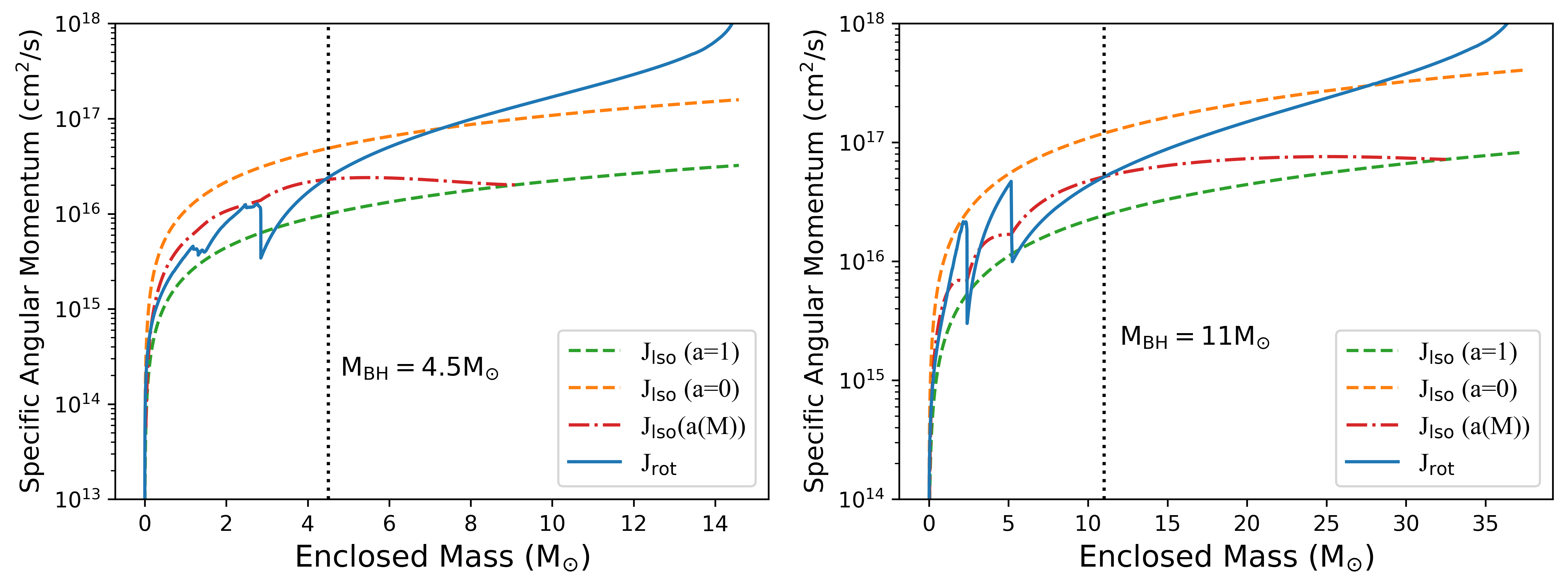}
    \caption{Specific angular momentum as a function of enclosed mass of our \textbf{(a)} Population I and \textbf{(b)} Population III progenitors. Blue lines show the rotational angular momentum of material at a given mass coordinate, while red dash-dotted lines show the angular momentum of a black hole formed from the collapse of all material within $M_{\mathrm{enc}}$. Green and orange dashed lines show the angular momentum of the inner most stable circular orbit of a maximally spinning and a non-rotating black hole, respectively. Where the blue line and red dash-dotted line cross marks the inner radial boundary of our RHD simulations.}
    \label{fig:jisco}
\end{figure*}

The choice of $r_{\mathrm{in}}$ described above can have dramatic implications on the probable physics of subsequent simulations. While stellar profiles obtained from  \texttt{MESA} include radii very close to the center of the star, the details of core-collapse in this region requires a treatment of 3D general relativistic magnetohydrodynamics, with still many open questions regarding the evolution of proto-neutron stars, and the formation of black holes and neutron stars. Here we assume the successful formation of an accreting black hole at the center of the star by cutting away the stellar profile within an inner boundary determined by the condition necessary for falling stellar material to form an accretion disk and power a jet. This condition requires that only stellar material with specific angular momentum $J_{\mathrm{rot}}>J_{\mathrm{lso}}$  can contribute to the jet. We therefore cut off the computational grid at $r_{\mathrm{in}}=r(M_{\mathrm{enc}})$, defined by $J(r_{\mathrm{in}})=J_{\mathrm{lso}}$, where
\begin{equation}
\label{eq:lso}
    J_{\mathrm{lso}} = \frac{G}{c} \frac{r_{\mathrm{lso}}^{2} M_{\mathrm{enc}}}{r_{\mathrm{lso}}^{3/2} + a(M_{\mathrm{enc}})M_{\mathrm{enc}}^{1/2}} ,
\end{equation}
 $a(M_{\mathrm{enc}})$ is determined from the relation
\begin{equation}
    \label{eq:angmom}
    J_{\mathrm{BH}} = a(M_{\mathrm{enc}}) \frac{GM_{\mathrm{enc}}^2}{c},
\end{equation}
and $r_{\mathrm{lso}}$ is the radius of the least stable orbit around a Kerr black hole \citep{bardeen_kerr_1970}. 
Figure \ref{fig:jisco} shows the specific angular momentum $J_{\mathrm{rot}}$ of both progenitors considered in this work, alongside the specific angular momentum of the last stable orbits of minimally and maximally spinning black holes as references, evaluated from Equation~\ref{eq:lso}.

\begin{figure}
    \centering
    \includegraphics[width=\columnwidth]{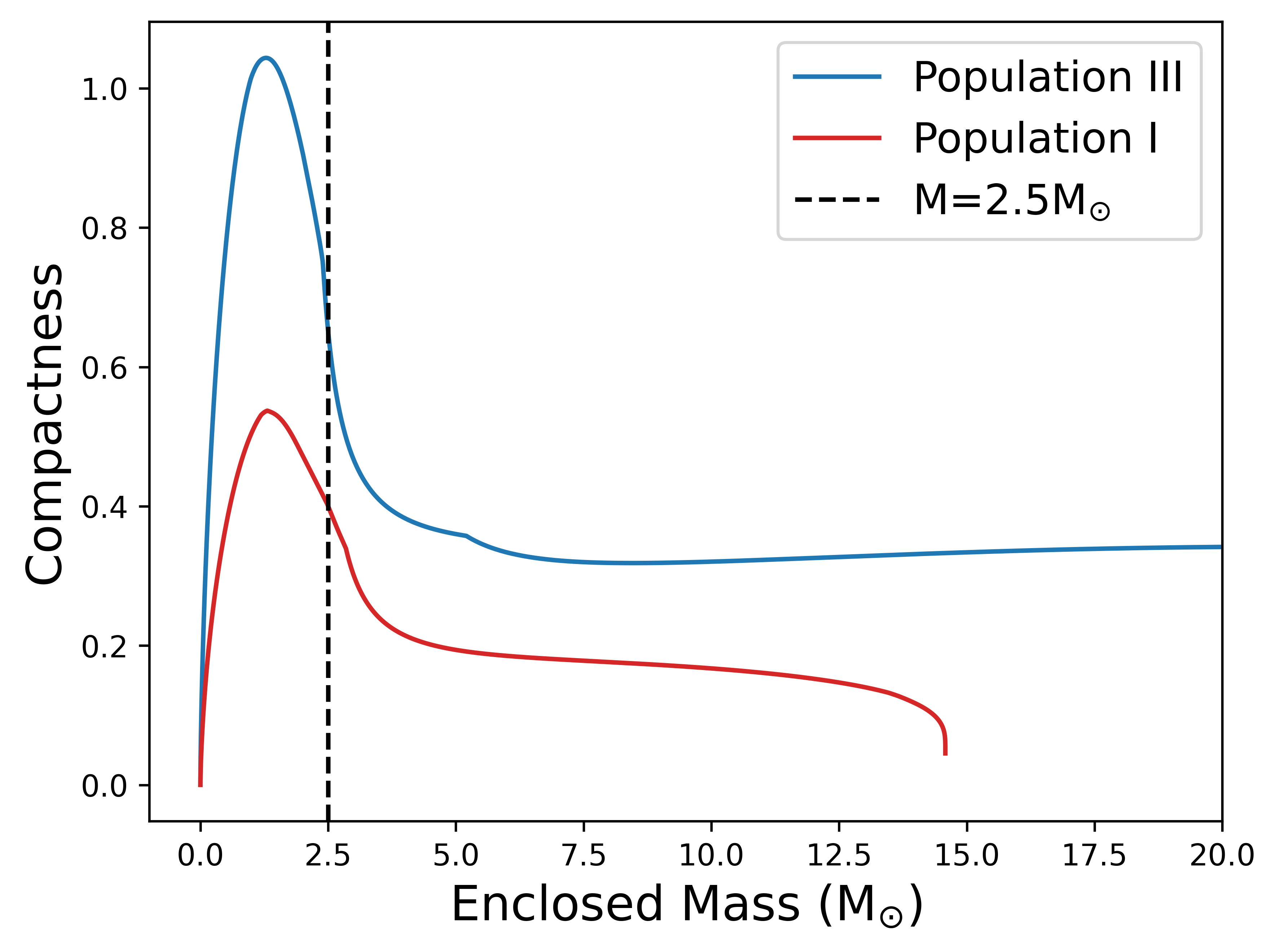}
    \caption{Compactness as a function of enclosed mass for both Population I and Population III progenitors. A compactness of $\gtrsim 0.4$ indicates a strong likelihood of a direct collapse to a black hole.  }
    \label{fig:compact}
\end{figure}

It should be noted that, while our stellar models are pre-collapse, we still use them as the initial conditions of a collapsar-driven GRB. We justify this through the estimation of the compactness of the progenitor, defined (e.g. \citealt{schneider_bimodal_2023})
\begin{equation}
    \xi_{M_{\mathrm{enc}}} = \frac{M_{\mathrm{enc}}/M_{\odot}}{R(M_{\mathrm{enc}})/1000 \, \mathrm{km}}\,.
\end{equation}
If the compactness at the core of the progenitor is high, a collapse to a black hole is likely. Alternatively, a low compactness favors an explosion scenario without the formation of a compact object. We show the compactness of both Population I and III models as a function of $M_{\mathrm{enc}}$ in Figure \ref{fig:compact}. In both cases, the compactness peaks for $M_{\mathrm{enc}} \lesssim 2.5 M_{\odot}$. Compactness between $0.4 - 0.8$ as seen here indicates the likely collapse of each stellar progenitor into a black hole.

\subsubsection{Jet Injection Procedure}

  As in \cite{nagakura_population_2012}, we calculate the jet luminosity as $L_{\mathrm{j}} = \eta\,\dot{M}\,c^2$, where $\dot{M} = \int_{\Sigma} \, da\,\rho v$ is the mass accretion rate across the surface $\Sigma$, and $\eta$ is the accretion efficiency. The simulations carried out in this work use 2D spherical coordinates, with accretion occurring through the inner radial boundary. The mass accretion rate then becomes
\newline
\begin{equation}
    \dot{M} = 2 \pi c \, r_{\mathrm{in}}^{2} \int_{\theta_j}^{\pi/2} d\theta \, \sin \theta \, \rho (r_{\mathrm{in}}, \theta) \, \beta^{r} (r_{\mathrm{in}},\theta)
    \label{eq:mdot}
\end{equation}

where $\theta_j$ is the half-opening angle of the jet, and $\rho=\rho(r_{\mathrm{in}},\theta)$ and $\beta^r = \beta^r(r_{\mathrm{in}},\theta)$ are the density and radial velocity, respectively, at the inner radial boundary of the domain, $r_{\mathrm{in}}$. While accretion efficiency has been shown to depend on the black hole spin \citep{lowell_rapid_2024} here we fix $\eta = 0.4$. The jet luminosity can then be related to hydrodynamical variables via $L_{\mathrm{j}} = \Delta \Omega \, c \, r^{2} \, \beta^{r} \, \Gamma (\Gamma_{\infty} - 1) \, \rho c^2$ (c.f. \citealt{walker_photospheric_2025}), where $ \Gamma_{\infty} = \Gamma \left(1+\frac{4p}{\rho c^2}\right)$, $\Gamma$ is the bulk Lorentz factor, $\rho$ is the density, $p$ is the pressure, and $\Delta \Omega = 2\pi \left( 1 - \cos \theta_j \right)$. Using Equation \ref{eq:mdot} to calculate $L_{j}$, and adopting $\theta_j=10^{\circ}$, $\Gamma = 10$ and $\Gamma_{\infty} = 400$, we then calculate $\rho$ and $p$.

\begin{figure}
    \centering
    \includegraphics[width=\columnwidth]{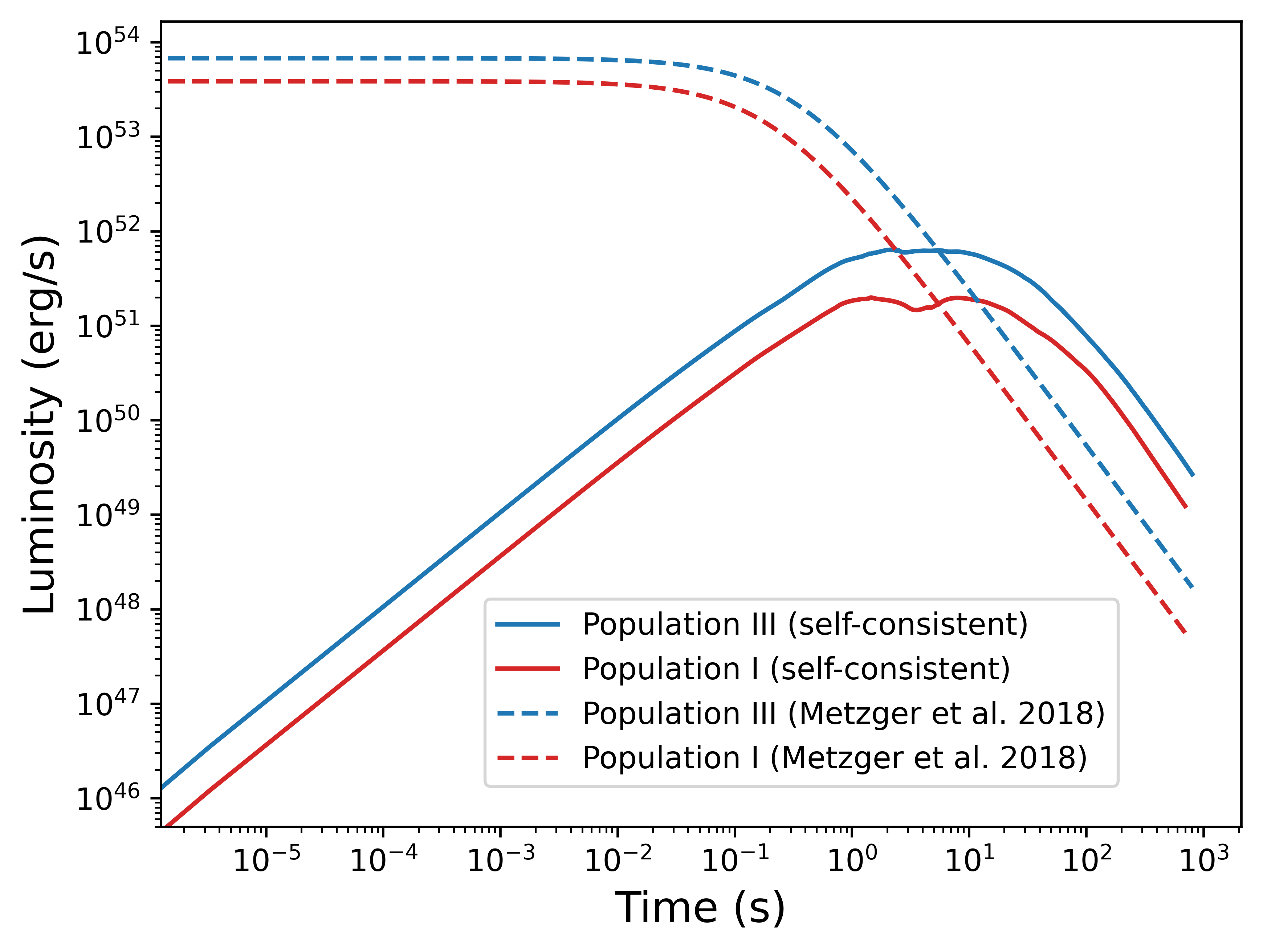}
    \caption{Jet luminosity as a function of time for both Population I (red) and Population III (blue) models. Dashed lines show the fallback luminosity of each model obtained via equation 1 in \cite{metzger_effects_2018}. Both self-consistent and fallback models approach the same power-law decay at late times, but otherwise differ quite dramatically.  }
    \label{fig:luminosity}
\end{figure}

Accretion is driven by the gravitational attraction of a black hole assumed to have a mass $M_{\mathrm{enc}}$ as determined by equations \ref{eq:lso} and \ref{eq:angmom}. In this way we calculate the jet injection energy according to the accretion rate at each timestep. This self-consistent, time-dependent jet luminosity is shown in Figure \ref{fig:luminosity} for both progenitors alongside corresponding analytical accretion rates given by\citep{metzger_effects_2018}
\begin{equation}
    \dot{M}(t) = \frac{2}{3} \frac{M_{\mathrm{fb}}}{t_{\mathrm{fb}}} \frac{1}{\left( 1 + t/t_{\mathrm{fb}}\,, \right)^{5/3}}
    \label{eq:fallback}
\end{equation}
where $M_{\mathrm{fb}}$ is the total fallback mass, $t_{\mathrm{fb}} \sim 1/\sqrt{G \bar{\rho}}$ is the characteristic fallback time, and $\bar{\rho}$ is the average density of the layer of the stellar progenitor contributing to accretion, here taken to be the density at the surface of the core taken from MESA simulations. We note that while both self-consistent and analytic accretion rates attain the same $t^{-5/3}$ power law at late times, their early time behavior differs greatly. Equation \ref{eq:fallback} is nearly constant for $t < t_{\mathrm{fb}}$, while Equation \ref{eq:mdot} results in $\sim t$ for $t < t_{\mathrm{fb}}$. This gradual increase for $t < t_{\mathrm{fb}}$ agrees with the self-consistent jet injection in \cite{nagakura_population_2012}, and is likely due to the pressure impeding the fallback of material, which is neglected in Equation \ref{eq:fallback}.

\subsection{Jet Structure and Evolution}

\begin{figure*}
    \centering
    \includegraphics[width=1.0\textwidth]{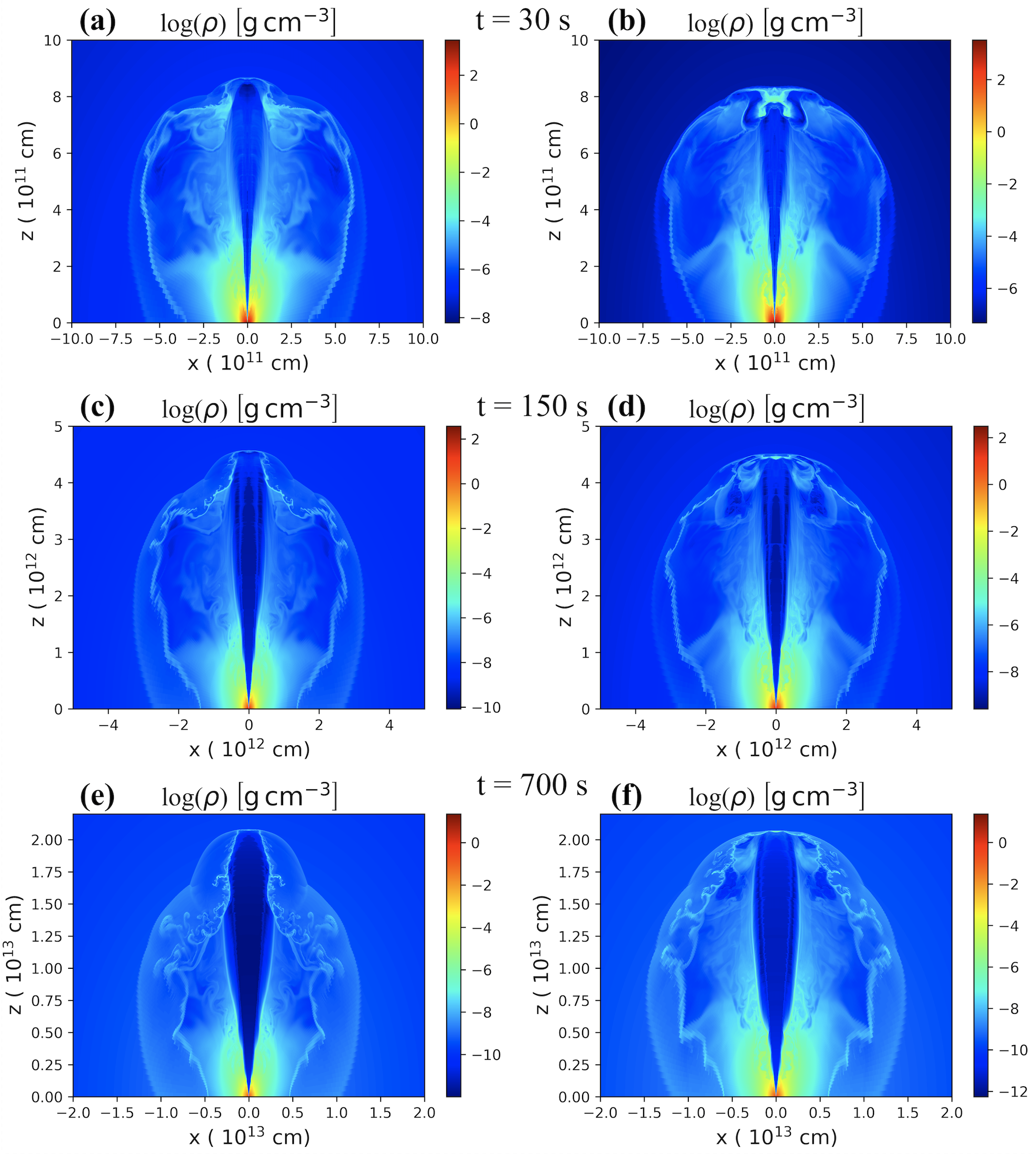}
    \caption{ Density maps of both Population I (left) and Population III (right) jets at $t\sim 30$ s (top row), $t \sim 150$ s (middle row), and $t \sim 700$ s. (bottom row). The Population III jet exhibits a plug at early times that becomes less noticeable later on. Both jets lack a clear "end", and have non-trivial angular and radial structure. }
    \label{fig:density_hydro}
\end{figure*}

We simulate jets from both progenitors for $\sim 700$ s, as shown in Figure \ref{fig:luminosity}. While the central engine is active for most of this time, peaks at $10^1 - 10^2$ provide a more reliable estimate of the engine time. It is during this interval in which the bulk of the energy is injected into the stellar progenitors. Figure \ref{fig:density_hydro} shows Population I and III jets at various times. At early times the Population III model exhibits a plug instability, which diminishes over time. The Population I model, in contrast, has no noticeable plug.

The structure of both jets display numerous signatures of a self-consistent luminosity, including the lack of clear radial boundaries of the jet outflow. As the jet luminosity never decays by more than a few orders of magnitude over a large fraction of the engine time, the jet nozzle reacts gradually to the material pushing back against the jet, which prevents it from being choked out completely. This effect is seen clearly near the base of the jet, which illustrates clearly intermittent periods where pressure from the environment temporarily overcame the jet pressure.

\begin{figure*}
    \centering
    \includegraphics[width=1.0\textwidth]{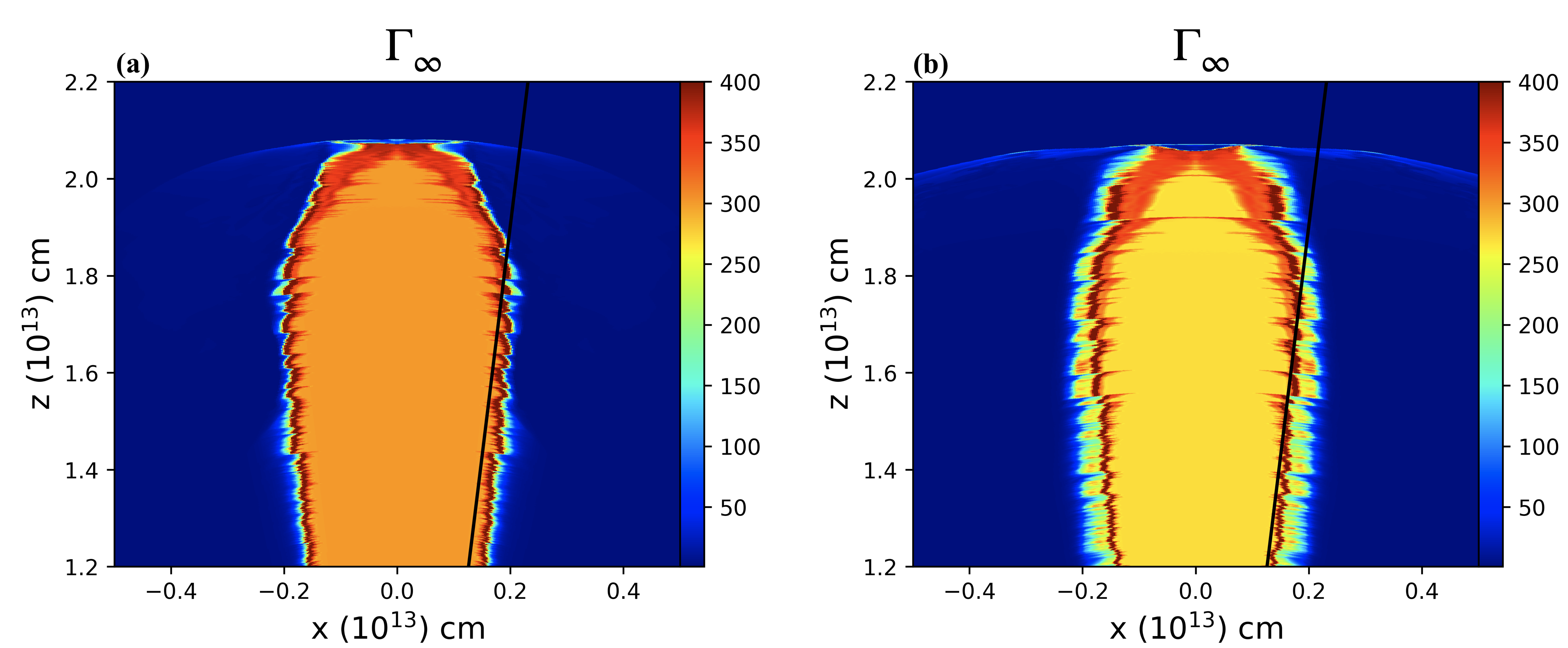}
    \caption{$\Gamma_{\infty}$ map of the final hydrodynamical frame for \textbf{(a)} our Population I and \textbf{(b)} our Population III models. The view presented here highlights the regions where the production of high-energy photons is likely to occur. Both jets have a region near the front of the jet head which can produce on-axis high energy photons. On-axis energy decreases at smaller radii, and the highest energy region is a 'shell' surrounding the core of the jet.  }
    \label{fig:gamma_inf_hydro}
\end{figure*}

\begin{figure*}
    \centering
    \includegraphics[width=1.0\textwidth]{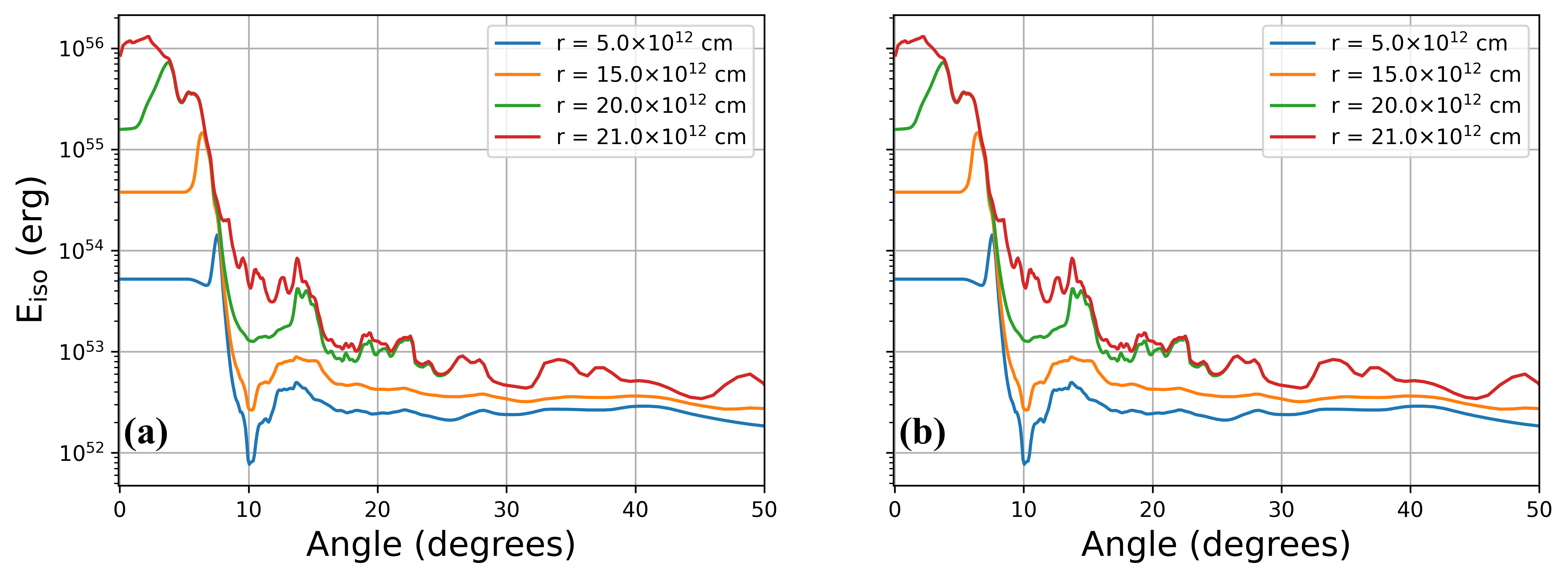}
    \caption{Isotropic equivalent energy as a function of angle, integrated along the line of sight relative to the jet axis for \textbf{(a)} our Population I and \textbf{(b)} our Population III models. Each line corresponds to the energy obtained from integrating over $r_{\mathrm{in}}$ to $r_{\mathrm{max}}$. For both Population I and III models $E_{\mathrm{iso}}$ peaks near the jet axis when close the the jet head. This peak is shifted to the right deeper inside the jet. }
    \label{fig:E_v_angle}
\end{figure*}

 Also evident is a non-trivial angular structure of the jet, which is shown in Figure \ref{fig:gamma_inf_hydro} with a zoomed-in perspective of the jet-head. Here we see a high-energy region near the front of the jet and close to the jet axis. However, further into the interior of both jets there is a relatively low energy core surrounded by a high energy shell that extends vertically down the sides of the jet envelope. There are also intermittent horizontal shocks, which in the case of the Population III model can cut all the way through the jet, whereas in the Population I model they only extend partially through the jet. In both cases the relatively low energy inner regions are mostly uniform.  This is also be seen in Figure \ref{fig:E_v_angle}, which shows the isotropic equivalent energy,

\begin{equation}
    E_{\mathrm{iso}} = 4 \pi \int_{r_{\mathrm{in}}}^{r_{\mathrm{max}}} \, dr \, r^2 \left( T^{00} - \Gamma \rho c^2 \right),
\end{equation}
where each line corresponds to a different value for $r_{\mathrm{max}}$. Here the jet energy peaks $\sim 2^{\circ}$ near the jet head. Further inside the jet, however, the peak is between $5^{\circ}$ and $10^{\circ}$. The radiation emitted by relativistic jets depends on the energy distribution of the jet along the line of sight. This structure can therefore impact the observed radiation, as will be seen in the next section.

% \begin{figure*}
%     \centering
%     \includegraphics[width=1.0\textwidth]{pressure_hydro.png}
%     \caption{Pressure map of the final hydrodynamical frame for \textbf{(a)} our Population III and \textbf{(b)} our Population I models.}
%     \label{fig:pressure_hydro}
% \end{figure*}

\section{Radiative Transfer}

We compute GRB prompt emission for both models using the post-processing radiation transfer code \texttt{MCRaT}  \citep{lazzati_monte_2016,parsotan_monte_2018}.\texttt{ MCRaT} injects photons into hydrodynamical output and Compton scatters them with electrons sampled from the hydrodynamical data. For further details on \texttt{MCRaT} we refer to existing literature.  

 We inject $\sim 10^7$ photons into both Population I and III simulations. The region of photon injection extends from $\theta = 0^{\circ}$ to $\theta = 6^{\circ}$ and lasts for 341 s.  We restrict photon injection to the first 6$^{\circ}$  to avoid issues stemming heavy wind profile we adopt in the hydrodynamical simulations. This wind has only a slight effect on on-axis emission, but can lead to an early external shock evident in off-axis emission. By considering photons only within the first $6^{\circ}$ of the jet axis we avoid this issue, while also considering regions of the jet with novel angular structure.

\subsection{Lightcurves and Spectra}

\begin{figure*}
    \centering
    \includegraphics[width=\textwidth]{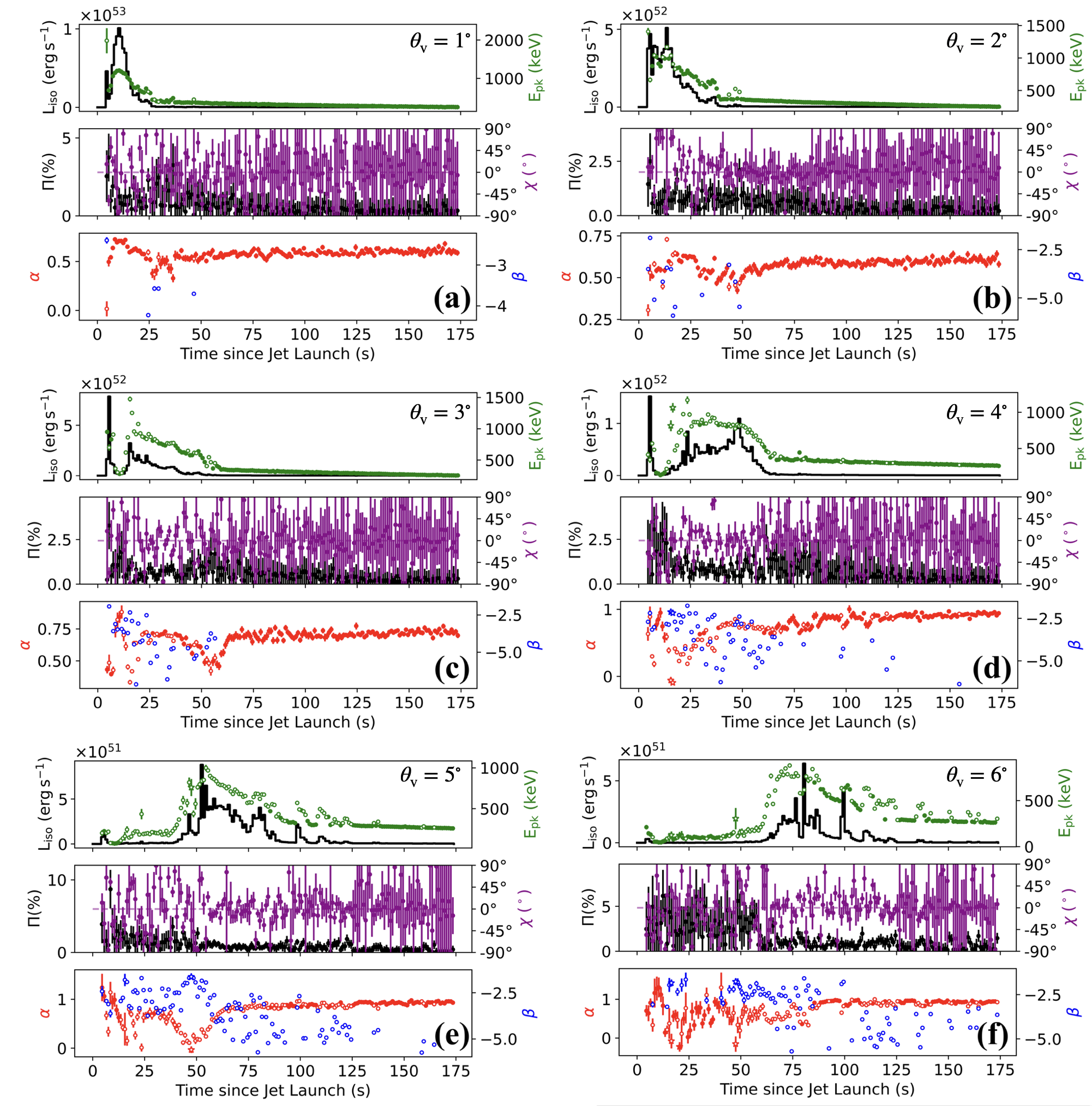}
    \caption{Lightcurves from our Population I model obtained from mock observations conducted at viewing angles $1^{\circ} - 6^{\circ}$ in 1$^{\circ}$ increments. In addition to lightcurves, each observation shows spectral fit parameters and the polarization degree and angle for each time bin.} 
    \label{fig:lightcurve_pop1}
\end{figure*}

Figure \ref{fig:lightcurve_pop1} shows lightcurves obtained from synthetic observations of Population I GRBs, conducted in the source frame, at viewing angles ranging from $1^{\circ}-6^{\circ}$.  Also shown are time-interval polarization and spectral parameters for each synthetic observation. We observe a minimum time delay between the jet launch and the detection of the first photon of $\sim 5$ s. This is constant with observing angles out to $\sim4^{\circ}$. Beyond this threshold lightcurves develop multiple peaks, with subsequent peaks dominating the first at large observing angles, which results in increased time delays.

\begin{figure*}
    \centering
    \includegraphics[width=\textwidth]{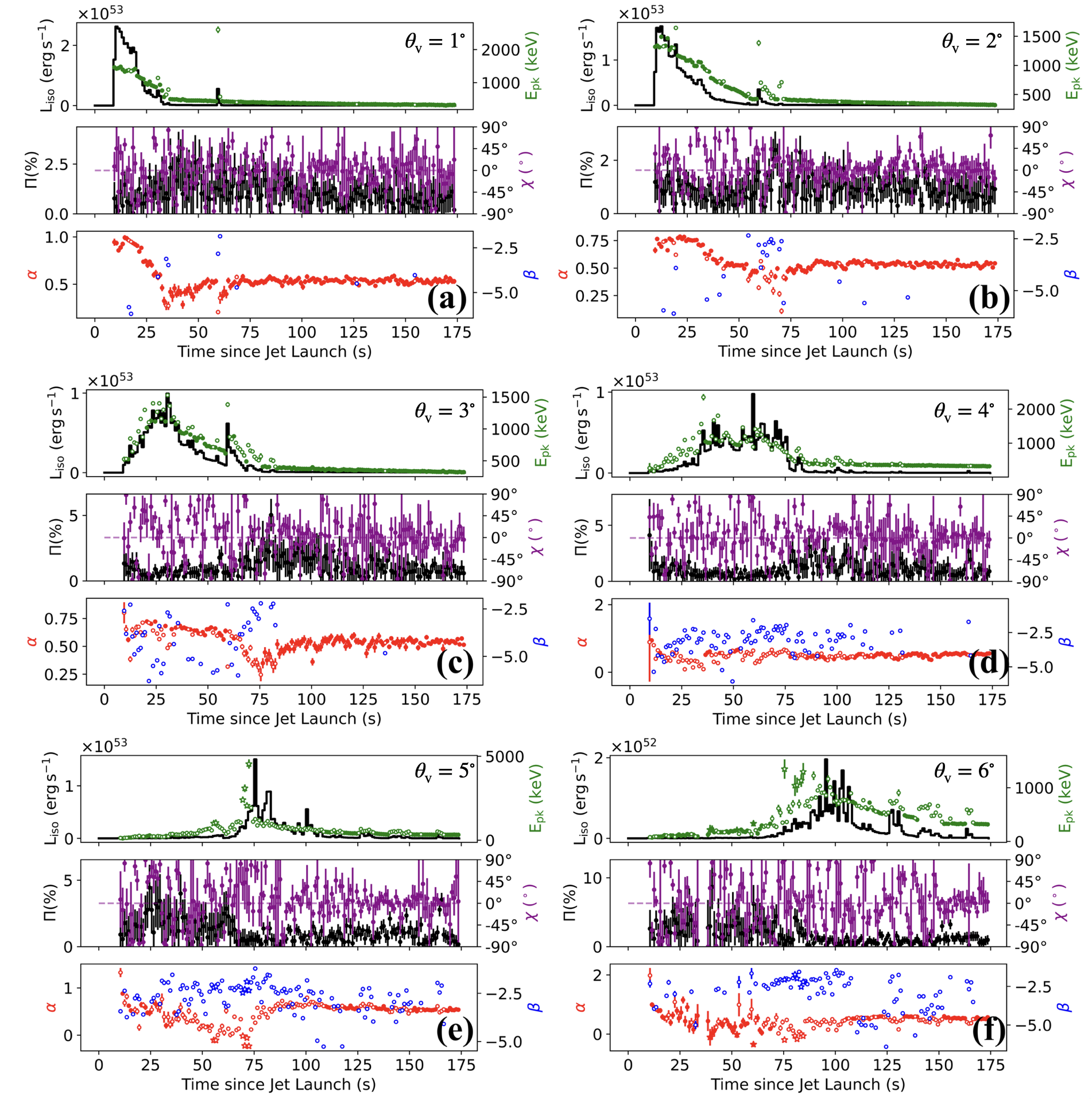}
    \caption{Same as Figure \ref{fig:lightcurve_pop1}, but for Population III model. Lightcurves here are brighter than their Population I counterparts by around a factor of 2, and generally have a single peak, with one notable exception, at $\theta{_\mathrm{v}}=3^{\circ}$. Population I and III lightcurves are otherwise similar.}
    \label{fig:lightcurve_pop3}
\end{figure*}

Figure \ref{fig:lightcurve_pop3} shows Population III lightcurves over the same range of viewing angles in Figure \ref{fig:lightcurve_pop1}. As with Population I lightcurves, at shallow angles we observe a mostly single peak burst that broadens as observing angle increases. Here, however, lightcurves largely have only one peak, except for at a viewing angle of $3^{\circ}$. Broadly speaking, Population III lightcurves are brighter and have a larger time delay of 10 s, compared to 5 s exhibited by Population I lightcurves.

\begin{figure*}
    \centering
    \includegraphics[width=\textwidth]{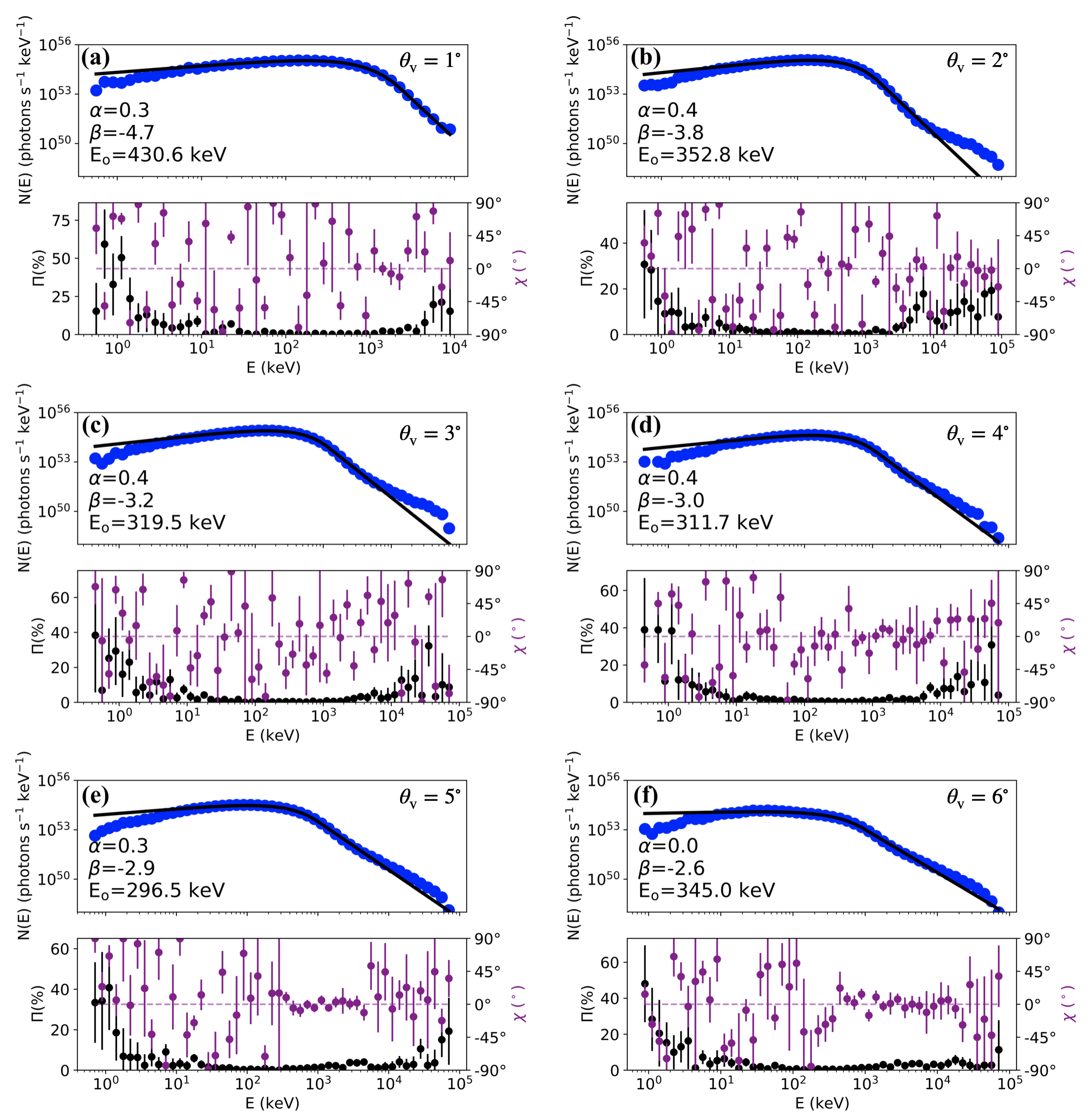}
    \caption{Spectra for our Population I model for the same viewing angles shown in Figures \ref{fig:lightcurve_pop1} and \ref{fig:lightcurve_pop3}, alongside time-integrated polarization parameters. On-axis observations are brighter and hotter, and become dimmer and colder off-axis, with the exception of viewing angle $\theta_{\mathrm{v}}=6^{\circ}$. We fit the Band spectrum \citep{band_batse_1993} to spectra obtained at each observing angle. } 
    \label{fig:spectrum_pop1}
\end{figure*}

\begin{figure*}
    \centering
    \includegraphics[width=\textwidth]{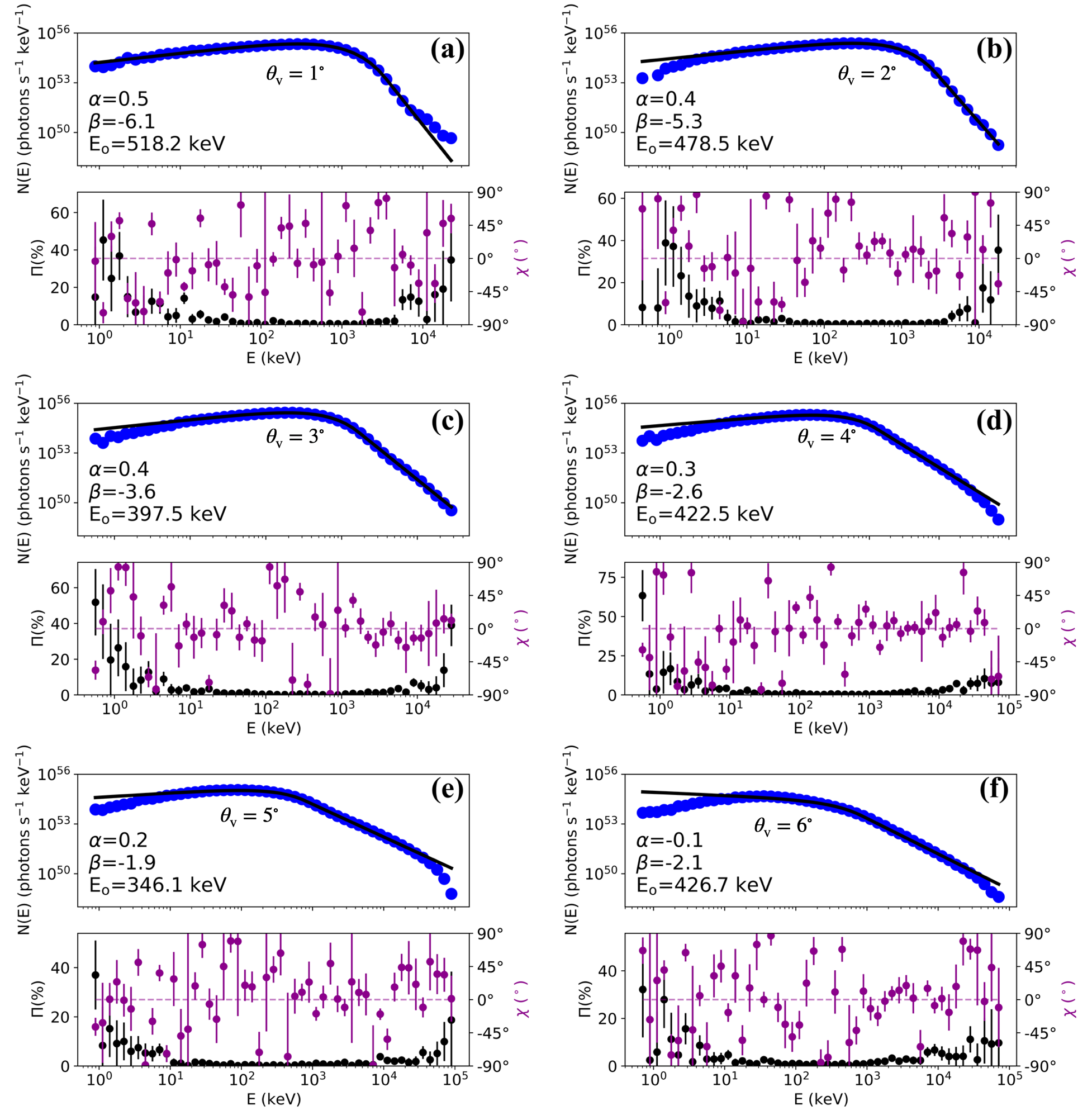}
    \caption{Same as Figure \ref{fig:spectrum_pop1} but for our Population III model. Population III spectra are generally hotter and brighter than their Population I counterparts.}
    \label{fig:spectrum_pop3}
\end{figure*}

Figures \ref{fig:spectrum_pop1} and \ref{fig:spectrum_pop3} show spectra corresponding to the synthetic observations shown in Figures \ref{fig:lightcurve_pop1} and \ref{fig:lightcurve_pop3}. As with lightcurves, Population III spectra are generally brighter than their Population I counterparts. Population III spectra also have higher peak frequencies, though not dramatically so. In both cases peak frequencies generally decrease with viewing angle, except for select observations with $\theta_{\mathrm{v}} \gtrsim 4^{\circ}$. The high-energy slope $\beta$ becomes increasingly non-thermal with viewing angle, while the low-energy slope $\alpha$ stays fairly thermal throughout, with the exception of observations at $\theta_{\mathrm{v}} = 6^{\circ}$for  both Population I and III models, which have  relatively non-thermal values for $\alpha$.

The degree of polarization is generally small across all observations. Values obtained from time resolved calculations (Figures \ref{fig:lightcurve_pop1} and \ref{fig:lightcurve_pop3}) are $\leq2\%$. There is some variation, with slightly lower polarization typically observed at times coinciding with lightcurve peaks. This effect is, however, often difficult to separate from errors. As expected, polarization angle is poorly constrained when the degree of polarization is so low. Time integrated polarization (Figures \ref{fig:spectrum_pop1} and \ref{fig:spectrum_pop3}) has more variability, with very high and very low energy photons being polarized up to $\sim 40\%$ and polarization angles exceeding $45^{\circ}$. The bulk of the observed photons, however, have very low polarization.

\begin{figure}
    \centering
    \includegraphics[width=\columnwidth]{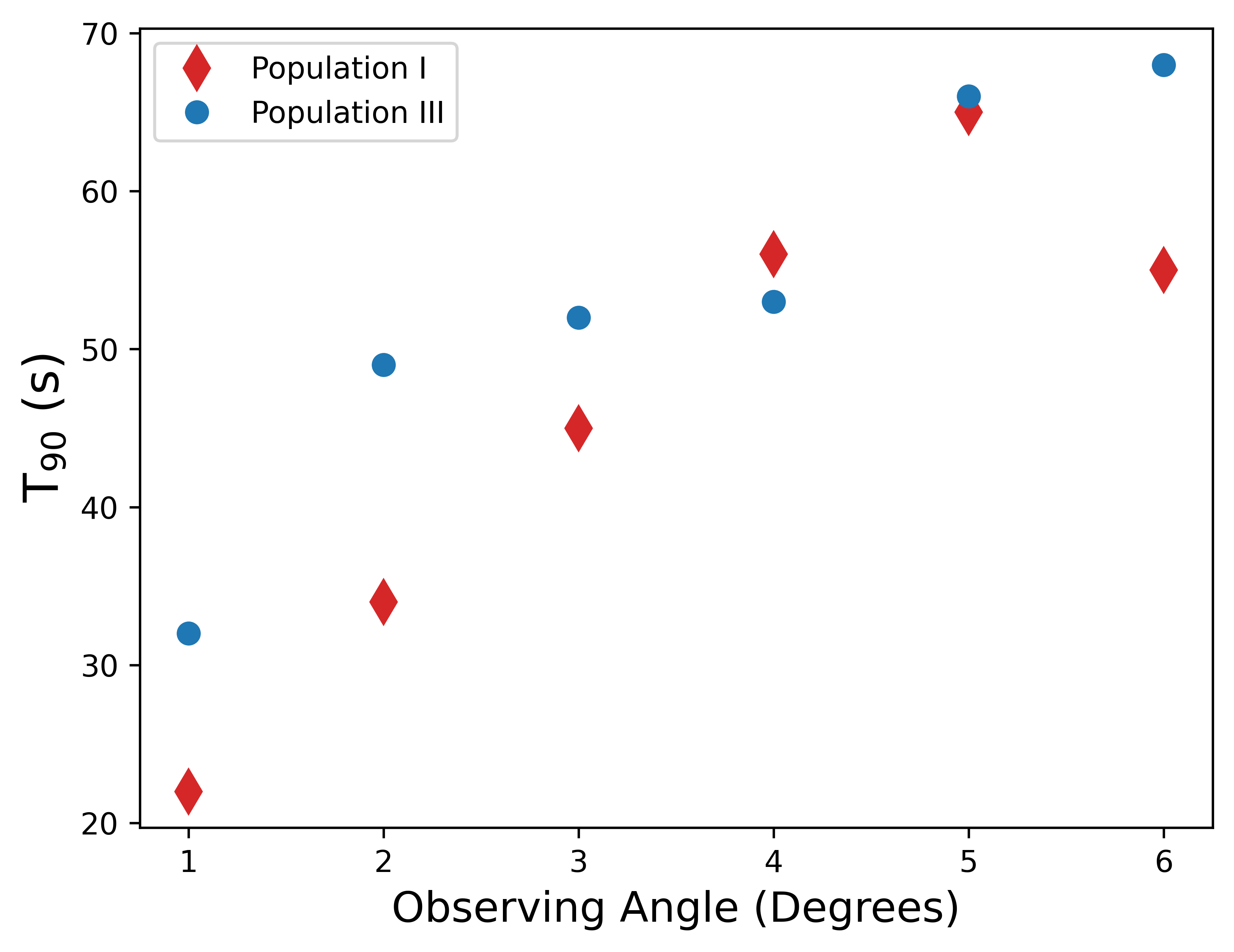}
    \caption{T90 as a function of viewing angle for the Population I (red) and Population III (blue) progenitors used in this work. Population III observations consistently have a longer $T_{90}$ than Population I observations.}
    \label{fig:T90}
\end{figure}

Figure \ref{fig:T90} shows $T_{90}$, obtained by integrating the lightcurves from Figures \ref{fig:lightcurve_pop1} and \ref{fig:lightcurve_pop3}. In both cases, $T_{90}$ increases steadily with viewing angle. Additionally, Population III observations have a longer $T_{90}$ than their Population I counterparts, with an exception at $\theta_{\mathrm{v}} = 4^{\circ}$.

\subsection{Observational Correlations}

\begin{figure*}
    \centering
    \includegraphics[width=\textwidth]{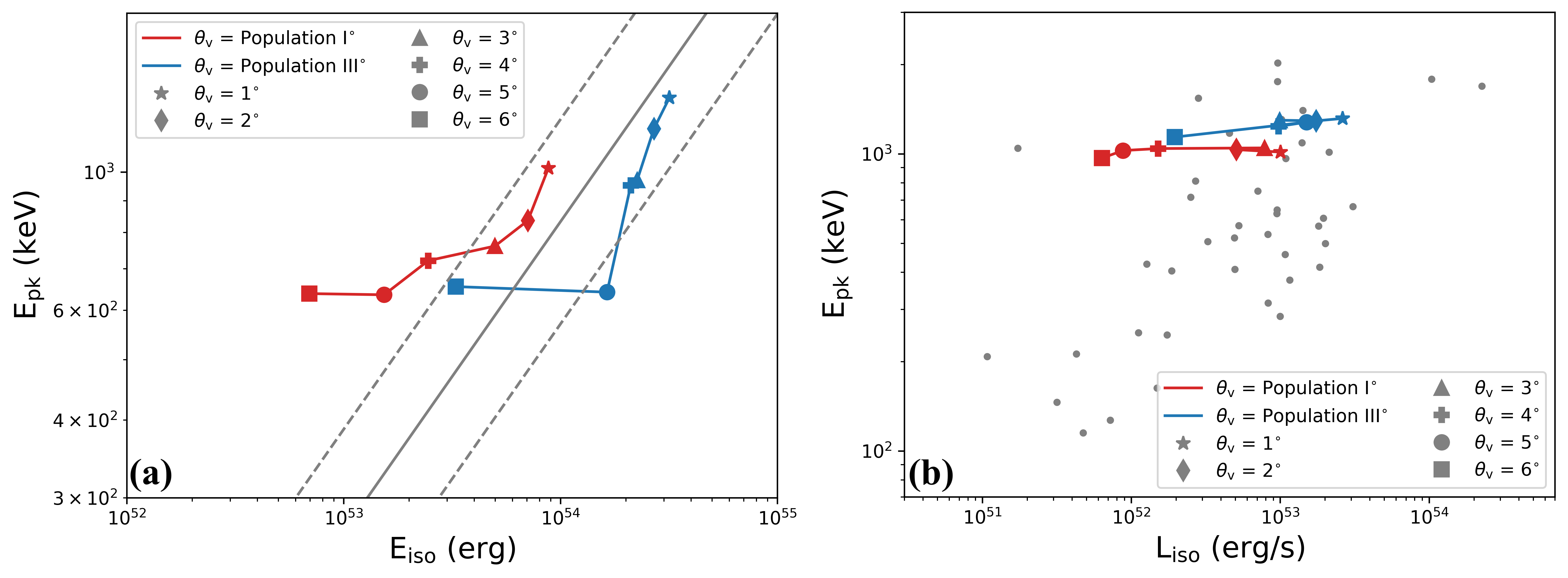}
    \caption{\textbf{(a)} The Amati \citep{amati_intrinsic_2002} and \textbf{(b)} Yonetoku correlations \citep{yonetoku_gamma-ray_2004} for both Population I and III models. We find good agreement with both relations. On-axis observations of both Population I and III bursts agree particularly well with the Amati relation. While our results don't agree particularly well with the slope of the Yonetoku relation, each synthetic observation does coincide with the set of real-world GRB observations. }
    \label{fig:amati_yonetoku}
\end{figure*}

Figure \ref{fig:amati_yonetoku} shows the Amati (panel \textbf{(a)}) and Yonetoku correlations (panel \textbf{(b)}) for Population I and III progenitors. In panel \textbf{(a)} we see agreement with the Amati relation for on-axis observations for both types of progenitors. Off-axis observations, however, strains this relation since the peak frequency does not decrease fast enough to compensate for the reduced isotropic energy. In panel \textbf{(b)} we find good agreement with the Yonetoku relation. It is worth noting that in a majority of cases Population I points lie upwards and to the left of Population III points. This is particularly clear with the Amati relation, where the majority of Population III observations lie below the relation, but within the 1$\sigma$ confidence interval, while the majority of Population I observations lie above the relation. 

\begin{figure*}
    \centering
    \includegraphics[width=\textwidth]{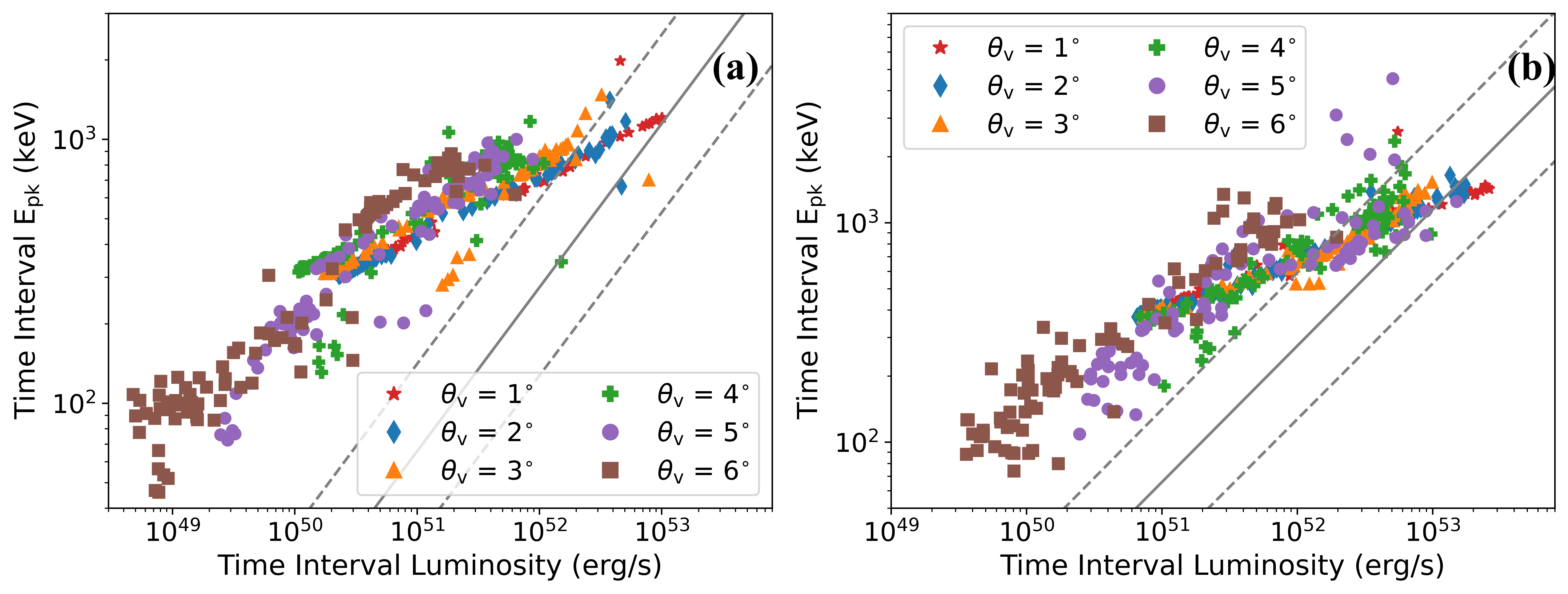}
    \caption{Time interval $E_{pk}$ plotted against time interval $L_{\mathrm{iso}}$, alongside the best-fit Golenetskii relation \citep{golenetskii_correlation_1983} for both \textbf{(a)} Population I and \textbf{(b)} Population III models. Each model shows observations obtained from the range of viewing angles shown in Figures \ref{fig:lightcurve_pop1} - \ref{fig:spectrum_pop3}. High-energy time-interval observations agree with this relation, while lower energy time-intervals exhibit significant strain. }
    \label{fig:golenetskii}
\end{figure*}

Figure \ref{fig:golenetskii} shows the Golenetskii correlation \citep{golenetskii_correlation_1983} for our MCRaT results for both Population I and III models, alongside the best fit Golenetskii relation. Both models strain this relation, with a large fraction of points lying significantly above the relation. Additionally, certain subsets of observations have a shallower slope than what is observed in nature.

\section{Summary and Discussion}
We have produced the first numerical results of GRB prompt emission resulting from the death of a Population III star, a task which required leveraging various technical advancements to treat the different phases of GRB production, including overcoming obstacles in the stellar evolution of zero-metallicity stars, as well as refining various aspects of the large-scale hydrodynamical and radiative processes that produce GRBs. 

We used \texttt{MESA} to simulate the stellar evolution of a rotating Population III star from zero-age main-sequence to just before core collapse. \texttt{MESA}, which treats angular momentum transport as a diffusive process in a standard 1D prescription, allows us to place constraints on the accretion power available to power a subsequent relativistic jet. 

Following calculations of stellar evolution, we conducted  a hydrodynamical simulation using the RHD module from \texttt{PLUTO}. In this simulation, we injected a time-dependent, self-consistent jet driven by the fallback of stellar material onto a compact remnant. Finally, we used the \texttt{MCRaT} radiation transfer code to scatter photons with the relativistic outflow and produce synthetic observations of the resulting GRB emission. Alongside this set of simulations based on a Population III stellar progenitor, we conducted analogous simulations using a reference Population I progenitor.

The use of a self-consistent jet luminosity results in dramatic changes to the overall jet structure and dynamics compared to past efforts using constant and analytic luminosities. Typical GRB jets (e.g. \citealt{lazzati_unifying_2012}) feature a fast, high energy core surrounded by a dense, mildly relativistic cocoon. The cores of these jets have significant radial structure \citep{morsony_origin_2010}, but little angular structure outside of the transition between the core and the cocoon. In contrast, here we observe dramatic angular and radial structure. The front of the jet is more typical of standard GRB jets. Beyond this, however, the core of the jet has a relatively low energy. Surrounding this core is a thin, high-energy shell which can produce off-axis emission. At even larger angles we observe a more traditional cocoon. We observe this structure in both Population I and III GRBs, which highlights the role that variable jet injection can have on the dynamics of GRB-producing jets, independent of progenitor.

Lightcurves from synthetic observations of both Population I and III GRBs have a similar shape.  While fine details do differ, it's difficult to attribute many of these to the difference in stellar progenitor. In both cases, lightcurves have peaks $\sim$ 20 s wide at shallow observing angles. At higher observing angles lightcurves are more diverse. Evidence of this includes significantly delayed bursts at observing angles $\gtrsim 5^{\circ}$. This delay is due to the aforementioned high-energy 'shell' surrounding both jets, which we identify as being due to the $\sim t^{-5/3}$ power law being prominent during a large fraction of the engine time. As the jet power slowly decreases, the cocoon continuously exerts pressure on the jet. This imparts a dramatic angular structure not typically seen in simulations of GRB jets.

Similarly, spectra from synthetic observations of Population I and III GRBs have many common features. We observe bright, hot spectra at shallow observing angles and colder, dimmer spectra at higher observing angles. There are exceptions to this for both progenitors around $\theta_{\mathrm{v}} \gtrsim 4^{\circ}$, which can have peak energies comparable to observations at $\sim2^{\circ}$. This feature arises from the off-axis high-energy shell described above.  We also find that the spectral high-energy slopes are relatively thermal at shallow observing angles, and become increasingly non-thermal at high observing angles. The spectral low-energy slopes, on the other hand, stay mostly thermal, except at $\theta_{\mathrm{v}}=6^{\circ}$ where $\alpha = 0, -0.1$ for Population I and III GRBs, respectively. 

We find promising agreement comparing these results to soft X-ray prompt emission observed by the Wide-field X-ray Telescope onboard the Einstein Probe. GRB 240315C \citep{liu_soft_2025}, which was jointly observed as a soft X-ray flash by the Einstein Probe and as a GRB by Swift-BAT and Konus-Wind, has a spectrum that agrees well with our Population III predictions. The peak energy of GRB 240315C falls within the range found here, and the Amati Correlation shown here (Figure \ref{fig:amati_yonetoku}) coincides with Type-II GRBs (see Figure 2 of \citet{liu_soft_2025}). Population III GRBs as sources for gamma-ray quiet-fast X-ray transients is less promising, however, as photon energies found in this work are too high to account for these \citep{angulo-valdez_ep260119a_2026}, even when accounting for cosmological redshift. We note, however, that prior MCRaT simulations of structured GRB jets \citep{parsotan_structured_2025} can explain both X-ray and gamma-ray observations due to viewing angle effects, which indicates the possibility that our procedure for predicting Population III GRBs can explain a wider range of X-ray transients, if extended to account for off-axis emission.

Both types of GRBs are also weakly polarized. Parameters from time-resolved calculations show photons are less than 5\% polarized across the entire lightcurves, with unconstrained polarization angles. Time integrated calculations, on the other hand, can yield polarizations as high as 40\% for photons that have either extremely high or extremely low energy. The bulk of these time-integrated photons have a polarization near 0\%, however.

While the above discussion suggests that it is difficult to distinguish between Population I and III GRBs, there are nevertheless features that can be used to discern between the two. In particular, Population III bursts are brighter than Population I bursts by about a factor of 2 for equal observing angles. Population III bursts are also hotter than their Population I counterparts. This is particularly noticeable in the Amati correlation, where  many Population I bursts lie above the relation, while Population III bursts lie below the relation. Consistent biases in the energetics of different bursts could provide a window into GRB progenitors. This effect is less noticeable in the Yonetoku and Golenetskii relations, though the shift in energies is still present. 

While we find agreement the Amati and Yonetoku relations, there is strain with the Golenetskii relation. Data points corresponding to the highest-energy portion of the lightcurve can lie withing the Golenetskii relation confidence interval, while points corresponding to low-energy regions often lie outside this interval. Taken as a whole, the slope is shallower than the real Golenetskii relation for both Population I and III bursts. 

Because points on the Golenetskii relation are obtained from photons coming from a localized region of the jet, we infer that certain regions of the jets presented in this work are capable of producing GRBs consistent with real-world observations, while other regions strain real-world observations. Furthermore, because the points on the Golenetskii relation which exhibit the most strain have relatively low energy, their contribution to the total isotropic energy is negligible. This explains the agreement we observe with the Amati relation. We also note that the total energy in observations is dependent on the total energy injected into our RHD simulations. Reducing the jet launching efficiency, among other parameters, would reduce the GRB isotropic energy. This could reduce the energy of the strained regions to the point where they don't produce detectable gamma-ray photons.

Another avenue for distinguishing between Population I and III GRBs is in the time delay between the jet launch and the detection of the first photons. This delay is, at minimum, due to the time it takes for the relativistic jet to break out of the stellar envelope. This is directly proportional to the mass of the progenitor, and this fact carries over to lightcurve time delays. Such a determination would, however, require observations of gravitational waves emitted during core collapse, a measurement which is well beyond current device sensitivities. This is also complicated by delayed bursts which arise from off-axis, making it difficult to extract an intrinsic time delay. We also note that Population III bursts have a consistently higher $T_{90}$ when compared to Population I bursts. 

Our results hint at a number of ways in which Population I and III GRBs may be distinguished, although it would be problematic to use any one of these features on its own. Combining several, along with potential measurements of observing angle and of the host galaxy, would allow greater flexibility in determining the progenitor of a given burst. Likewise, it is of considerable importance to consider the scope of our results and asses how resilient they will be in future studies of GRB progenitors.

In this regard, the approximations inherent to 1D stellar evolution can in principle affect GRB emission. While the methods we adopt for incorporating 3D effects into 1D stellar evolution calculations are widely used, they rely on simplified parameterizations calibrated primarily against observations of nearby, metal-rich stars. Recent observational and theoretical studies have highlighted significant uncertainties in the treatment of rotational mixing and angular momentum transport, suggesting that additional processes, such as wave-driven transport, may operate in stellar interiors \citep[e.g.,][]{Cantiello2014, Rogers2013, Aerts2019}. Moreover, rotating stellar models continue to face challenges in reproducing the observed surface abundance patterns of massive stars \citep[e.g.,][]{Hunter2008, Dufton2018}, with further complications arising from binary interactions and magnetic effects \citep[e.g.,][]{Langer2012, deMink2013, Meynet2011}. Consequently, predictions of stellar structure for rapidly rotating massive stars, particularly in the metal-free regime are largely exploratory. However, we find that the clearest effects that a stellar progenitor can have on GRBs depend largely on the mass of the progenitor. While finer details of stellar structure and rotation may have a noticeable impact, these changes will not likely contradict the results presented here.

Beyond approximations associated with 1D stellar evolution, the parameters chosen in this work are in some respects favorable to the successful launch of a relativistic jet. Jet launching efficiencies of $40\%$ are nearly at the theoretical maximum for the Blandford-Znajek mechanism \citep{blandford_electromagnetic_1977}, and are clearly favorable for extracting enough energy from a collapsing star to power a jet. However, increasing jet power also increases the possibility of feedback strong enough to terminate accretion, which would provide an obstacle to jet launching. Here we demonstrate the ability for collapsar-powered accretion to sustain a long-lived relativistic jet even at high efficiencies. There is still the consideration that reducing the jet power while keeping the mass of the stellar envelop constant may impede the jet. We note that this is not likely to be a problem, as the jets considered here launch quickly after the simulation begins and do not show signs of being choked out. We  therefore conclude that jets can be launched successfully over a wide range of possible jet efficiencies. 

Details of GRBs that arise from self-consistent injection may still be sensitive to parameters in other respects than whether or not a jet successfully launches. Investigating this possibility, however, would require a broad exploration of parameter space for both the stellar progenitors the jet injection procedure. Given the numerous ways in which we find that stellar progenitors and the physics jet launching can impact observed emission, and the promising agreement we find with observations carried out by the Einstein Probe, this will line of inquiry has the potential to play a crucial role in connecting GRB observations to their progenitors. Further study can therefore not only  enable more accurate discrimination between Population I and III GRBs, but also has the potential to shed light on stellar remnants, GRB central engines, and accretion physics.

\section*{Acknowledgements}

D.L. and N.W. gratefully acknowledge support by NASA award 80NSSC25K7554.
The Center for Computational Astrophysics at the Flatiron Institute is supported by the Simons Foundation. High-end computing (HEC) resources supporting this work were provided by the NASA Advanced Supercomputing (NAS) Division at Ames Research Center.

\software{\texttt{MESA} \citep{paxton_modules_2010,paxton_modules_2013,paxton_modules_2015,paxton_modules_2018,paxton_modules_2019},
\texttt{PLUTO} \citep{mignone_pluto_2007}
\texttt{MCRaT} \citep{lazzati_monte_2016,parsotan_monte_2018},
\texttt{Matplotlib} \citep{hunter_matplotlib_2007},
\texttt{ProcessMCRaT} \citep{parsotan_parsotatprocessmcrat_2021}.}
\FloatBarrier
\bibliography{bib}{}
\bibliographystyle{aasjournal}

\end{document}